\documentclass[twocolumn,aps,preprintnumbers,prl,showpacs,amsfonts,amsmath,amssymb,superscriptaddress,floatfix]{revtex4}

\usepackage[dvips]{graphicx}
\usepackage[tight]{subfigure}

\allowdisplaybreaks[3]

\newcommand{\hatd}[1]{\hat{#1}^\dagger}

\newcommand{\avg}[1]{\langle{#1}\rangle}
\newcommand{\bra}[1]{\langle{#1}|}
\newcommand{\ket}[1]{ |{#1} \rangle}

\newcommand{\bracket}[2]{\langle{#1}|{#2}\rangle}
\newcommand{\bradmket}[3]{\langle #1 | #2 | #3 \rangle}

\newcommand{\In}{\mathrm{in}}
\newcommand{\Out}{\mathrm{out}}
\newcommand{\A}{\mathrm{A}}
\newcommand{\B}{\mathrm{B}}
\newcommand{\C}{\mathrm{C}}

\newcommand{\affA}{%
    Department of Applied Physics and Quantum-Phase Electronics center, School of Engineering,
        The University of Tokyo,\\
    7-3-1 Hongo, Bunkyo-ku, Tokyo 113-8656, Japan}

\newcommand{\affB}{%
    Optical Quantum Information Theory Group,
Max Planck Institute for the Science of Light,\\
Institute of Theoretical Physics I, Universit\"{a}t Erlangen-N\"{u}rnberg,
Staudtstr.7/B2, 91058 Erlangen, Germany}

\begin{document}

\title{Demonstration of a universal one-way quantum quadratic phase gate}

\date{\today}

\author{Yoshichika Miwa}
\affiliation{\affA}
\author{Jun-ichi Yoshikawa}
\affiliation{\affA}
\author{Peter van Loock}
\affiliation{\affB}
\author{Akira Furusawa}
\affiliation{\affA}


\begin{abstract}
We demonstrate a quadratic phase gate
for one-way quantum computation in the continuous-variable regime.
This canonical gate, together with phase-space displacements
and Fourier rotations, completes the set of universal gates
for realizing any single-mode Gaussian transformation such as
arbitrary squeezing. As opposed to previous implementations of
measurement-based squeezers, the current gate is fully controlled
by the local oscillator phase of the homodyne detector.
Verifying this controllability, we give an experimental
demonstration of the principles of one-way quantum computation
over continuous variables. Moreover, we can observe sub-shot-noise
quadrature variances in the output states, confirming that
nonclassical states are created through cluster computation.

\end{abstract}


\maketitle

Measurement-based one-way quantum computation~\cite{Raussendorf01.prl},
using an offline prepared, multi-party entangled cluster state,
is a conceptually interesting alternative to the
standard unitary circuit model of quantum computation~\cite{Nielsen2000}.
In the cluster-model, universality is achieved through
different choices of measurement bases, while the cluster state
remains fixed. Unitary gates are effectively applied at each measurement
step, corresponding to elementary teleportations~\cite{Zhou00.pra,Nielsen2006}
for propagating and manipulating a quantum state through the cluster.
The cluster model also turned out to provide new,
potentially more efficient approaches to the experimental
realization of quantum logical gates, especially in the
quantum optical setting~\cite{Nielsen2004,Browne2005}.

A translation of the circuit model for quantum computation
over continuous variables (CV)~\cite{lloyd,SamPvLRMP}  to universal cluster
computation with CV was given in ~Ref.~\cite{clusterMenicucci}. The canonical,
universal gate set for CV is $\{\hat U_3(\lambda),C\}$, where
$C=\{\hat Z(s),\hat U_2(\kappa),\hat F,C_Z\}$ with the momentum shift operator
$\hat Z(s) = \exp(2 i s \hat x)$, the phase gates
$\hat U_l(\kappa_l) = \exp(i\kappa_l \hat x^l)$, the Fourier transform operator
$\hat F$, and the controlled-$Z$ gate
$C_Z = \exp(2i \hat x\otimes\hat x)$~\cite{Bartlett}.
Through concatenation, the full set enables one to simulate any
Hamiltonian in terms of arbitrary polynomials of the position $\hat x$ and
the momentum $\hat p$ to any precision~\cite{lloyd}.

The same set without the cubic gate $\hat U_3$, i.e., the set $C$,
is still universal for realizing any quadratic Hamiltonian,
that is, the whole group of Gaussian unitary transformations, the analogue
to the Clifford group for discrete variables (DV).
In the case of DV, for example, single-qubit Clifford transformations
are fully covered by the Hadamard gate $\hat H$ and the
`$\pi/4$'-phase gate $\hat U_{\pi/4}$ acting upon the qubit Pauli operators as
$\hatd U_{\pi/4}Z\hat U_{\pi/4}= Z$ and
$\hatd U_{\pi/4}X\hat U_{\pi/4}=-i X Z=-Y$; full universality
for single-qubit transformations would then require, in addition, the
well-known `$\pi/8$'-phase gate~\cite{Nielsen2000}, the analogue to the
cubic phase gate $\hat U_3$ for CV. Focussing on CV, the quadratic gate
from the universal set $C$ for all Gaussian transformations, maps the
Weyl-Heisenberg displacement operators $\hat Z(s)$ and
$\hat X(s)=\exp(-2 i s \hat p)$ into
\begin{align}
&\hatd U_2(\kappa) \hat Z(s) \hat U_2(\kappa) = \hat Z(s) \nonumber\\
&\hatd U_2(\kappa) \hat X(s) \hat U_2(\kappa) = e^{-i \kappa s^2} \hat X(s) \hat Z(-\kappa s),
\end{align}
in analogy to the qubit `$\pi/4$'-phase gate $\hat U_{\pi/4}$.
The effect of the phase gate may be more conveniently expressed
in terms of the generators of the Weyl-Heisenberg group,
$\hatd U_2(\kappa) \hat x \hat U_2(\kappa) = \hat x$,
$\hatd U_2(\kappa) \hat p \hat U_2(\kappa) = \hat p + \kappa \hat x$.

In quantum optics, it is well-know that there is an exact,
finite decomposition of any quadratic unitary into single-mode squeezers
and beam splitters~\cite{Braunstein05.pra,Reck}.
In this quantum optical language, the quadratic phase gate
$\hat U_2$, together with the Fourier transform $\hat F$,
provides single-mode squeezing, and the two-mode gate
$C_Z$ involves beam splitting modulo single-mode squeezing.

In the cluster-based one-way model,
the quadratic gate can be fully controlled through the local oscillator
phase of the homodyne detector~\cite{clusterMenicucci}. Here,
we experimentally demonstrate this controllability,
with a fixed, offline two-mode cluster state.
We show that a large set of squeezing transformations can be achieved
by means of this one-way phase gate; sequential application of the gate
would lead to universal single-mode Gaussian transformations
(where changes of the 1st moments in phase space require, in addition,
$p$-displacements $\hat Z(s)$, trivially realizable through
a cluster state for CV~\cite{clusterMenicucci}).

The output states
of our elementary cluster computations exhibit sub-shot-noise
quadrature variance; thus, nonclassical states are created
deterministically through cluster computation with the
degree of nonclassicality fully
controlled by the measurement apparatus.
Therefore, our demonstration differs from previous implementations of
universal offline squeezing~\cite{Filip05.pra,Yoshikawa07.pra},
in which different squeezing transformations require
different beam splitter transformations to achieve universality.







The elementary teleportation step for the
case of CV~\cite{clusterMenicucci} is described as follows.
First, in the ideal scheme (Fig.~1(a)),
an arbitrary input state is coupled
to a single-mode, infinitely squeezed state (a position eigenstate $\ket{x = 0}$),
$\hat{U}_{\mathrm{QND}} \ket{\psi}_\In\ket{x=0}_{\A}$. This results in
$e^{-2i\hat{x}_\In\hat{p}_\A} \int dx\,   \psi (x)
\ket{x}_\In \int dp\,\ket{p}_{\A}/\sqrt{\pi}
= \int dx \psi(x)\ket{x}_{\In}\ket{x}_{\A}$,
where the subscripts `$\In$' and `$\A$' denote the input and ancilla modes, respectively.
Up to local Fourier rotations, the resulting state corresponds to a perfect two-mode
cluster state, already carrying the quantum information to be processed through
the cluster (i.e., the quantum state $\ket{\psi}_\In$).

Next, we measure the observable $\hatd{U}(\hat{x})\hat{p}\hat{U}(\hat{x})$ of mode 1, where
$\hat{U}(\hat{x})\equiv \exp[i f(\hat x)]$ is diagonal in the position basis and
$\hat{p}$ is the conjugate momentum to $\hat{x}$ ($[\hat{x},\hat{p}]=i/2$).
The quantum state after the measurement with outcome $p_0$ is
$\sqrt{\pi}\,{_\In\!\bradmket{p_0}{\hat{U}(\hat{x}_\In ) \int \psi(x)}{x}_\In}\ket{x}_{\A} dx
= \sqrt{\pi} \int {_\In\!\bracket{p_0}{x}_\In} U(x) \psi (x) \ket{x}_{\A} dx
= \hat{Z}(-p_0)\hat{U}(\hat{x}_\A )\ket{\psi}_\A$.
After correcting the displacement $\hat{Z}(-p_0)$,
we obtain the desired state $\hat{U}(\hat{x})\ket{\psi}$ in the ancilla mode.
Through this scheme, in principle, we can apply an arbitrary unitary
operator $\hat{U}(\hat{x})$ to $\ket{\psi}_\In$;
for nonlinear gates such as the cubic gate $\hat U_3$, however, this would require
measuring a nonlinear observable. Here, we consider detection of the whole range of rotated
quadratures (all linear combinations of $\hat x$ and $\hat p$), effectively applying the
quadratic phase gate $\hat U_2(\kappa)=\exp\left(i\kappa\hat{x}^2\right)$
to $\ket{\psi}_\In$, up to a phase-space displacement
depending on the measurement result $p_0$.




In our optical realization, $\hat{x}$ and $\hat{p}$ are quadrature operators,
for the mode operator $\hat{a}=\hat{x}+i\hat{p}$. The quadratic gate $\hat U_2(\kappa)$ corresponds to a sequence of rotation,
squeezing, and rotation~\cite{Braunstein05.pra}, with  $\hat{x}_\Out = \hat{x}_\In$ and
$\hat{p}_\Out = \hat{p}_\In +\kappa \hat{x}_\In$.
Thus, the required measurement corresponds to measuring~\cite{clusterMenicucci}
$\hat{p} + \kappa \hat{x}=
\sqrt{1+\tan^2 \theta}\left(\hat{p}\cos \theta + \hat{x}\sin \theta \right)$ with
$\kappa = \tan\theta$.
Using homodyne detection and setting the phase of the local oscillator (LO) to $\theta$, we can measure $(\hat{p}\cos\theta + \hat{x}\sin\theta)$.
Appropriate electric amplification of the homodyne results with gain $(1+\tan^2 \theta)^{1/2}$ leads to the desired measurement of $\hatd{U}\hat{p}\hat{U}$.
We show this for several values of $\kappa$: $0$, $\pm 1.0$, $\pm 1.5$, $\pm 2.0$, with coherent-state inputs.
The corresponding LO phases are $0^\circ$ , $\pm 45^\circ$, $\pm 56.3^\circ$, and $\pm 63.4^\circ $, respectively.

\begin{figure}[tb]
\centering
\subfigure[Schematic of elementary one-way quantum computation.]{
\includegraphics[clip, scale=0.4]{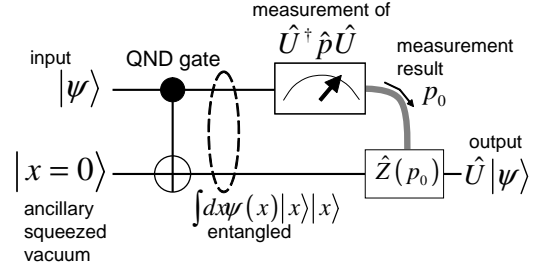}
}
\subfigure[Our implementation of one-way quantum computation.]{
\includegraphics[clip, scale=0.33]{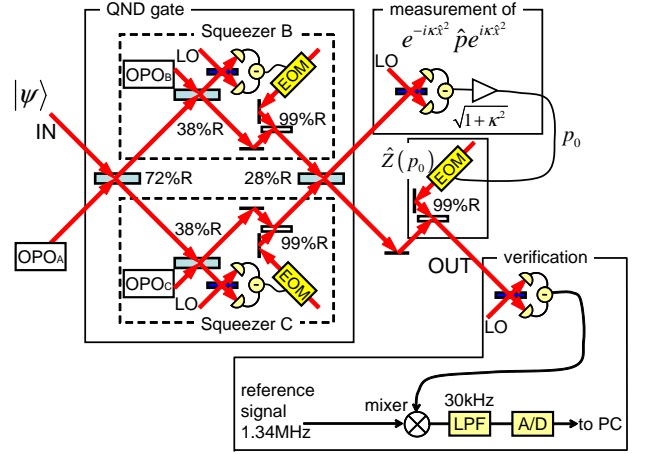}
}
\caption{Schematic of a one-way quantum gate and our experimental setup. 
OPO: optical parametric oscillator, LO: optical local oscillator, and EOM: electro-optic modulator.}
\label{fig:whole}
\end{figure}

In our optical demonstration, we use three squeezed-vacuum ancillae.
One ancilla is coupled to the input via a QND gate (denoted by subscript $\A$). The QND gate itself requires two additional squeezed vacuum states (denoted by subscripts $\B$, $\C$). For the QND gate, we employ the scheme of Refs.~\cite{Filip05.pra,Yoshikawa08.prl}. 
The full input-output relations of the scheme including finite-squeezing resources are
\begin{align}
\hat{x}_\Out &= \hat{x}_\In +\hat{x}_\A^{(0)}e^{-r_\A} -\frac{\sqrt{5}-1}{2\sqrt[4]{5}}\hat{x}_\B^{(0)}e^{-r_\B}, \notag \\
\hat{p}_\Out &= \hat{p}_\In +\kappa \hat{x}_{\In} + \frac{1}{\sqrt[4]{5}}\kappa \hat{x}_\B^{(0)}e^{-r_\B} + \frac{\sqrt{5}+1}{2\sqrt[4]{5}}\hat{p}_\C^{(0)}e^{-r_\C}.
\label{eq:input-output with excess noise}
\end{align}
Even with the excess noise from the finite squeezing of the ancillae,
we are able to observe sub-shot-noise quadrature squeezing for sufficiently large $\kappa$.
In the remainder of the letter, we shall describe the experimental details and present the results of the experiment.






{\it Experimental setup.}---
A schematic of the experimental setup is illustrated in Fig.~1(b).
The original source of light is a continuous wave (CW) Ti:sapphire laser, whose output is 860~nm in wavelength and 1.5~W in power.
Quantum states at the 1.34 MHz sideband are used in our demonstration.

The experimental setup consists of the following parts: preparation of the input and ancilla states, the QND coupling gate, measurement, feedforward, and, finally, the verification measurement.

The input state, a coherent state at the 1.34~MHz sideband, is generated by modulating a weak laser beam of about 10~$\mu$W using electro-optic modulators (EOMs).
We prepare three types of coherent states $\ket{\alpha}$:
$\alpha = x_\In$, $\alpha = i p_\In$, and $\alpha = 0$, corresponding to phase modulation, amplitude modulation, and no modulation of the laser beam, respectively.

In order to prepare the ancilla states, there are three sub-threshold optical parametric oscillators (OPOs), each generating a single-mode squeezed state, whose squeezing level is $-4.3$dB, $-4.9$dB, and $-5.2$dB.
An OPO is a bow-tie shaped cavity of 500~mm in length, containing a PPKTP crystal~\cite{Suzuki06.apl}.
The second harmonic (430~nm in wavelength) of Ti:sapphire output is divided into three beams in order to pump the OPOs.


The QND gate basically consists of a Mach-Zehnder interferometer with a single-mode squeezing gate in each arm~\cite{Yoshikawa08.prl}. Each single-mode squeezing gate contains a squeezed vacuum ancilla, homodyne detection, and feedforward~\cite{Filip05.pra, Yoshikawa07.pra}.
The variable beam splitters in the QND gate are composed of two polarizing beam splitters and a half-wave plate.
We can eliminate the QND gate and just measure the input states by setting the transmittances of the variable beam splitters to unity.
At each beam splitter, we lock the relative phase of the two input beams by means of active feedback to a piezoelectric transducer.
For this purpose, two modulation sidebands of 154~kHz and 107~kHz are used as phase references.
For the homodyne detection, the LO phase is adjusted in accordance to the desired $\kappa$ value; the feedforward
displacement is carried out with the right gain depending on $\kappa$.

To verify the output state, we employ another homodyne detection.
As is well known from optical homodyne tomography, we can reconstruct the quantum state from the marginal distributions for various phases~\cite{Lvovsky09.rmp}.
We slowly scan through the LO phase and perform a series of homodyne measurements.
The 1.34 MHz component of the homodyne signal is extracted by means of lock-in detection: it is mixed with a reference signal and then sent through a 30~kHz low pass filter.
Finally, it is analog-to-digital converted where the sampling rate is 300,000 samples per second.

The powers of the LOs are about 3~mW.
The detector's quantum efficiencies are greater than 99\%, the interference visibilities to the LOs are on average 98\%, and the dark noise of each homodyne detector is about 17~dB below the optical shot noise level produced by the LO.
Propagation losses of our whole setup are about 7\%.

\begin{figure}[tb]
\centering
\subfigure[Input coherent state.]{
\includegraphics[clip,scale=0.38]{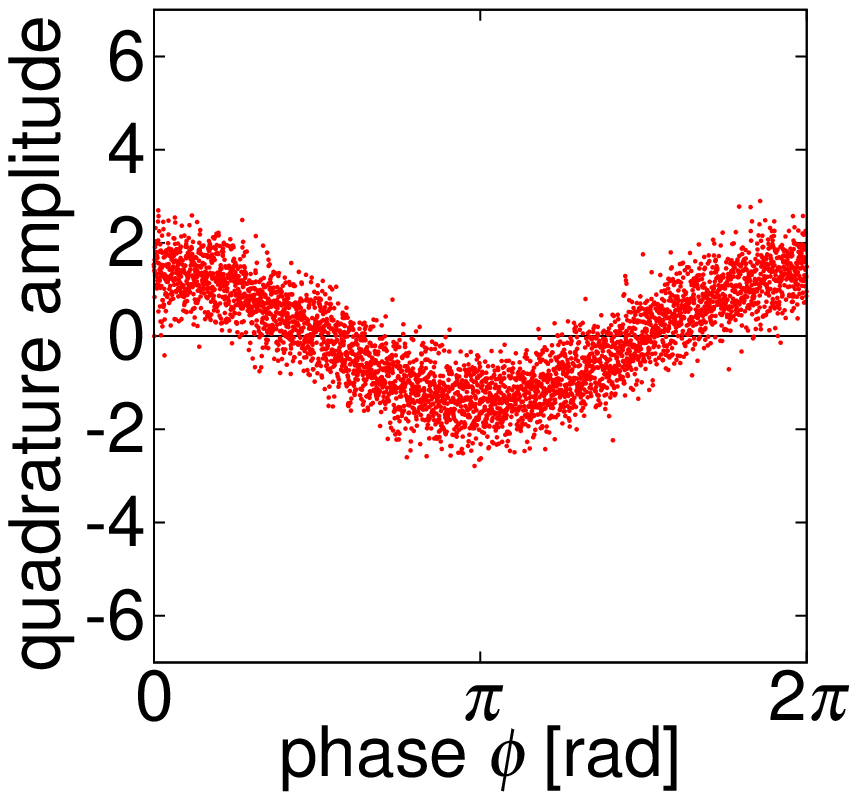}
\includegraphics[clip,scale=0.5]{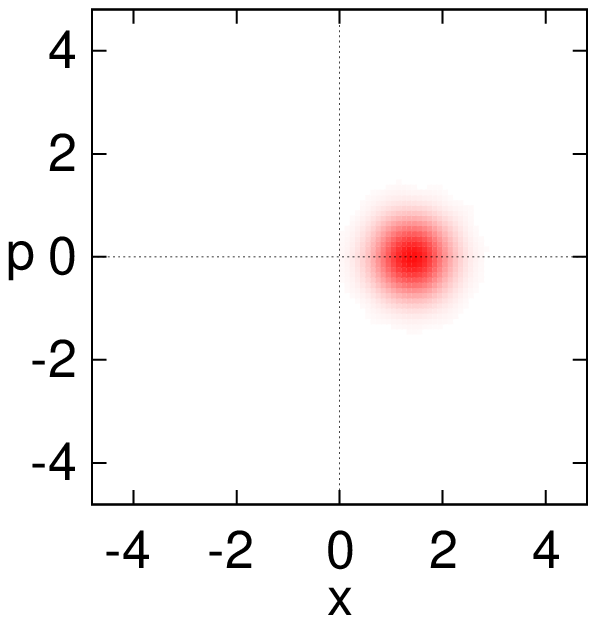}
}
\subfigure[Output for $\kappa=0$.]{
\includegraphics[clip,scale=0.38]{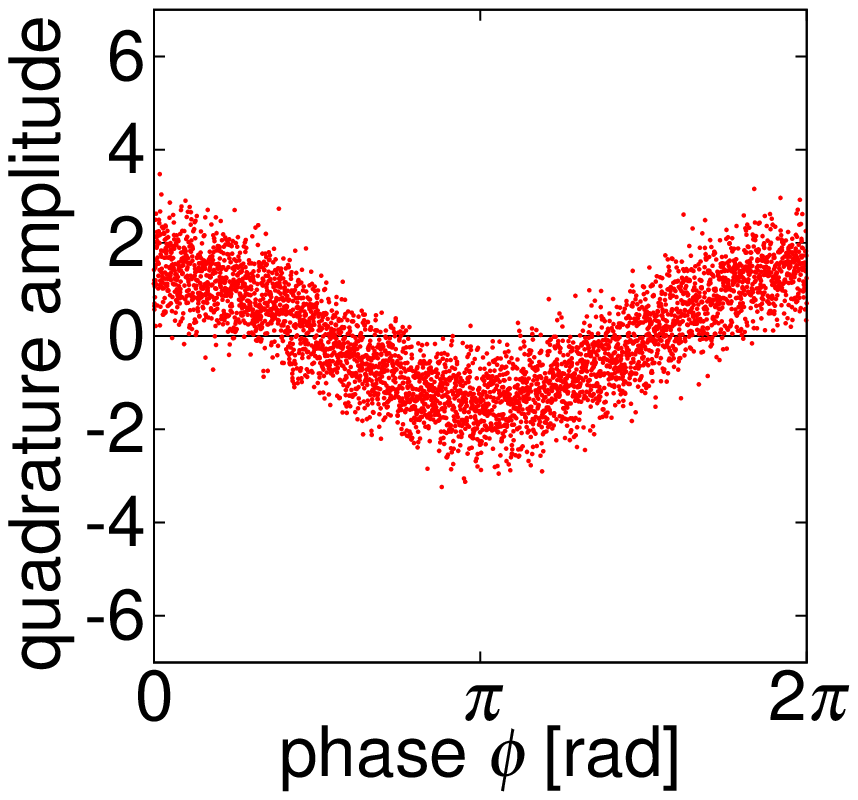}
\includegraphics[clip,scale=0.5]{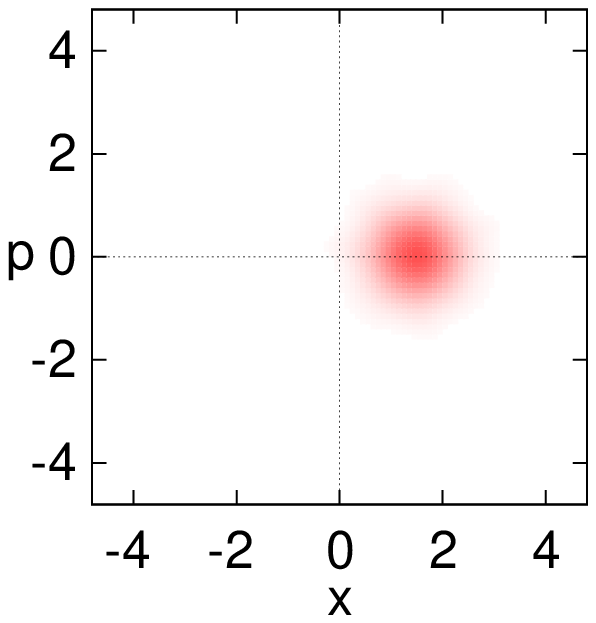}
}
\subfigure[Output for $\kappa=1.0$.]{
\includegraphics[clip,scale=0.38]{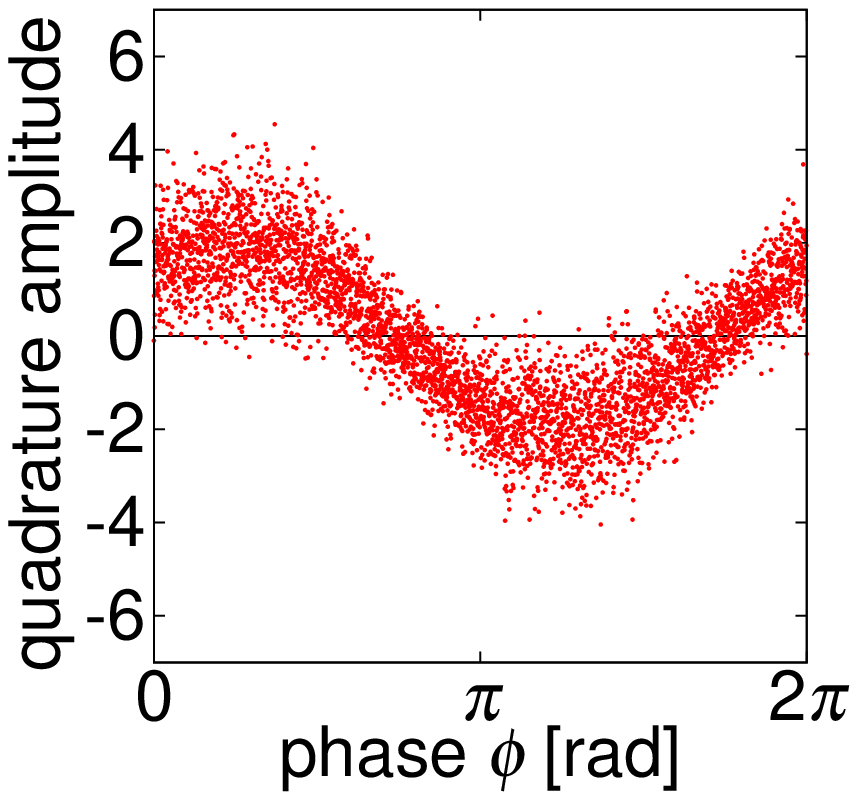}
\includegraphics[clip,scale=0.5]{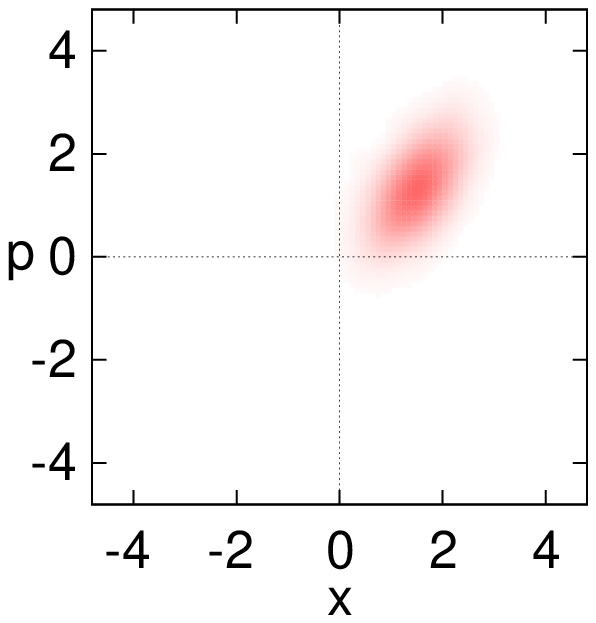}
}
\subfigure[Output for $\kappa=2.0$.]{
\includegraphics[clip,scale=0.38]{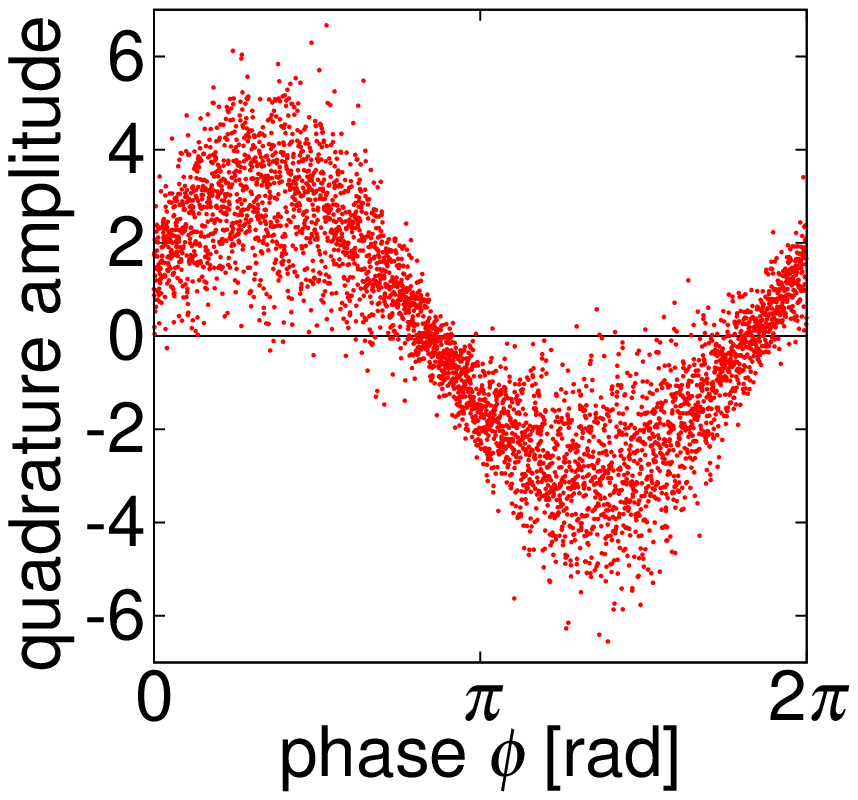}
\includegraphics[clip,scale=0.5]{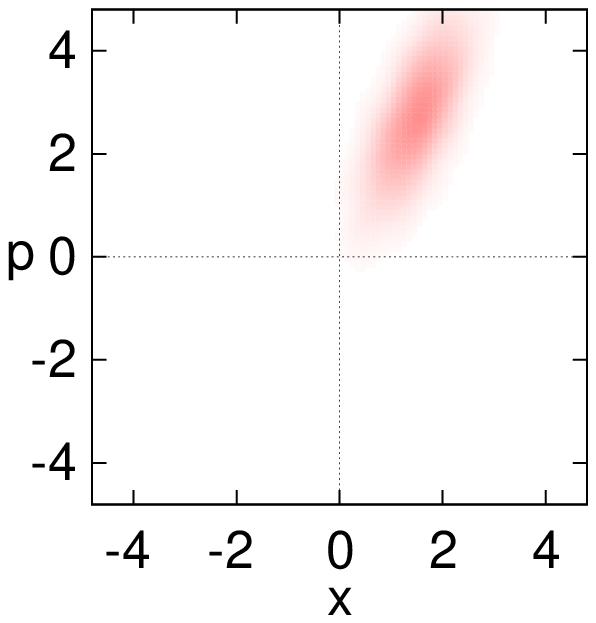}
}
\caption{Input and output states with several $\kappa$. Left figures show raw data of marginal distributions and right ones show the Wigner functions, reconstructed via maximum-likelihood method~\cite{Lvovsky04}.}
\label{fig:results}
\end{figure}

{\it Experimental results.}---
As mentioned earlier, we carry out the experiment with three types of input coherent states $\ket{\alpha}$:
$\alpha = x_\In$ ($x_\In = 1.4$), $\alpha = i p_\In$ ($p_\In = 1.3$), and $\alpha = 0$.
For each input state, we demonstrate the gate for seven different $\kappa$ values: $0$, $\pm 1.0$, $\pm 1.5$, and $\pm 2.0$.

Fig.~2 shows the raw data of marginal distributions and the Wigner functions reconstructed via maximum-likelihood method~\cite{Lvovsky04}.
We show the results for the input state with the amplitude in $x$ as an example.
Each scan contains about 80,000 data points which are uniformly distributed in phase from 0 to $2\pi$, and every 20 points are plotted in the figure (about 4,000 data points).
For $\kappa = 0$ (Fig.~2(b)), the input state is regenerated at the output except for some excess noise.
For nonzero $\kappa$ (Fig.~2(c, d)), we can see that the distribution of the $p$ variable is shifted proportional to $x$, with a proportionality factor $\kappa$. As a result, the output states are squeezed and rotated.


\begin{figure}[tb]
\centering
\subfigure[$\alpha = x_\In$ ($x_\In = 1.4$).]{
\includegraphics[clip,scale=0.45]{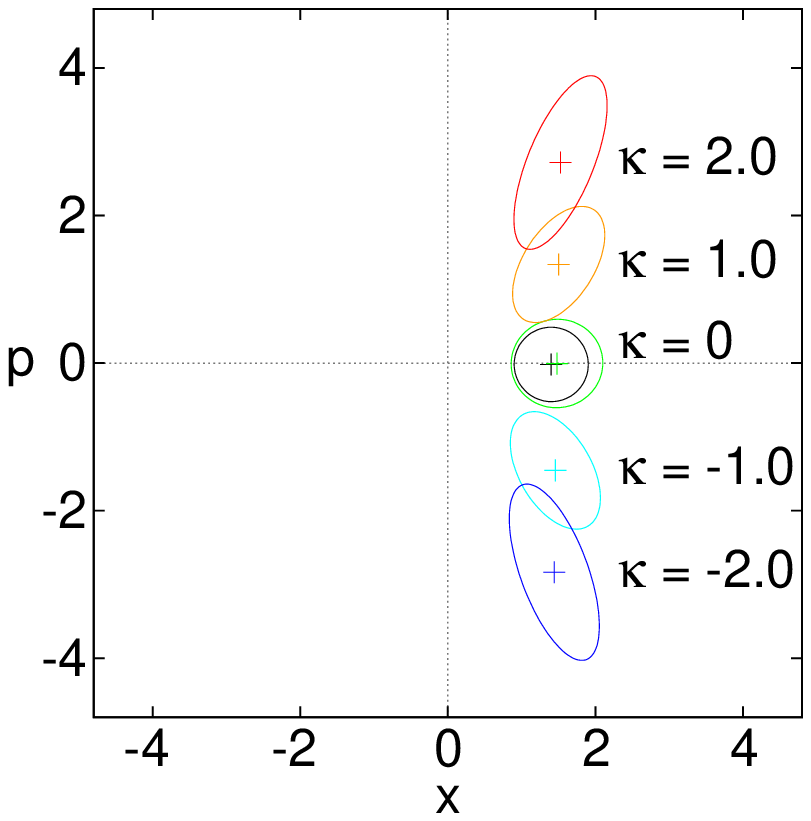}
}
\subfigure[$\alpha = x_\In$ (ideal operation).]{
\includegraphics[clip,scale=0.45]{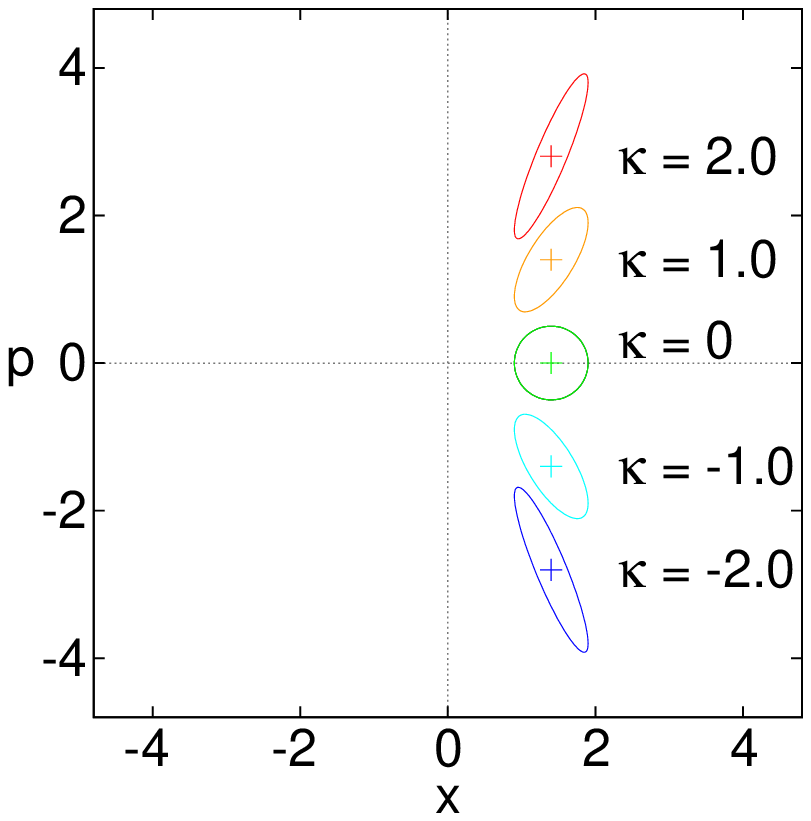}
}
\subfigure[$\alpha = 0$]{
\includegraphics[clip,scale=0.45]{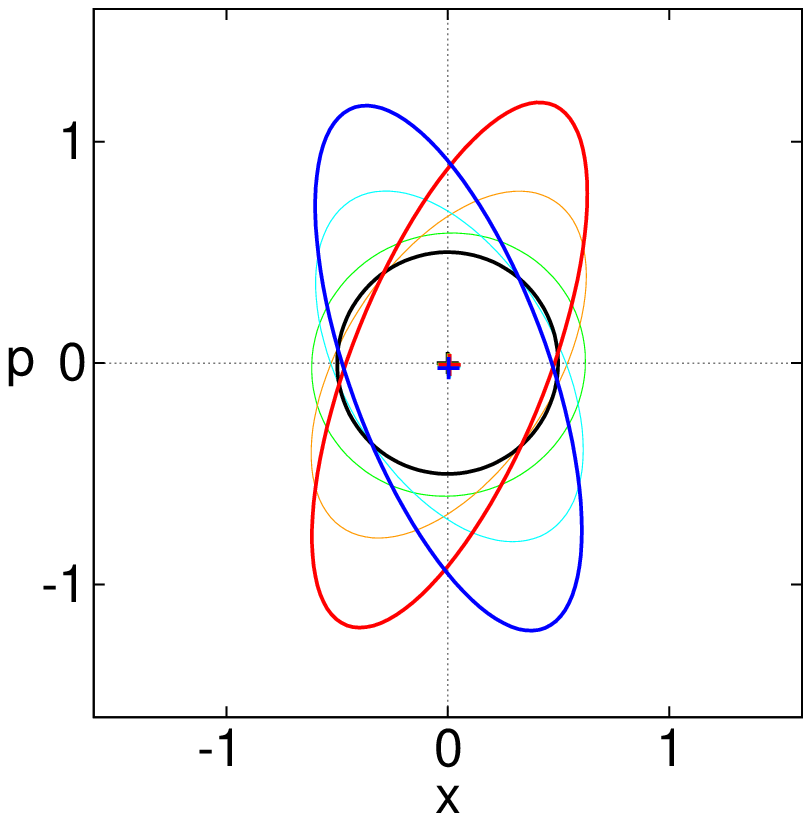}
}
\subfigure[$\alpha = ip_\In$ ($p_\In = 1.3$).]{
\includegraphics[clip,scale=0.45]{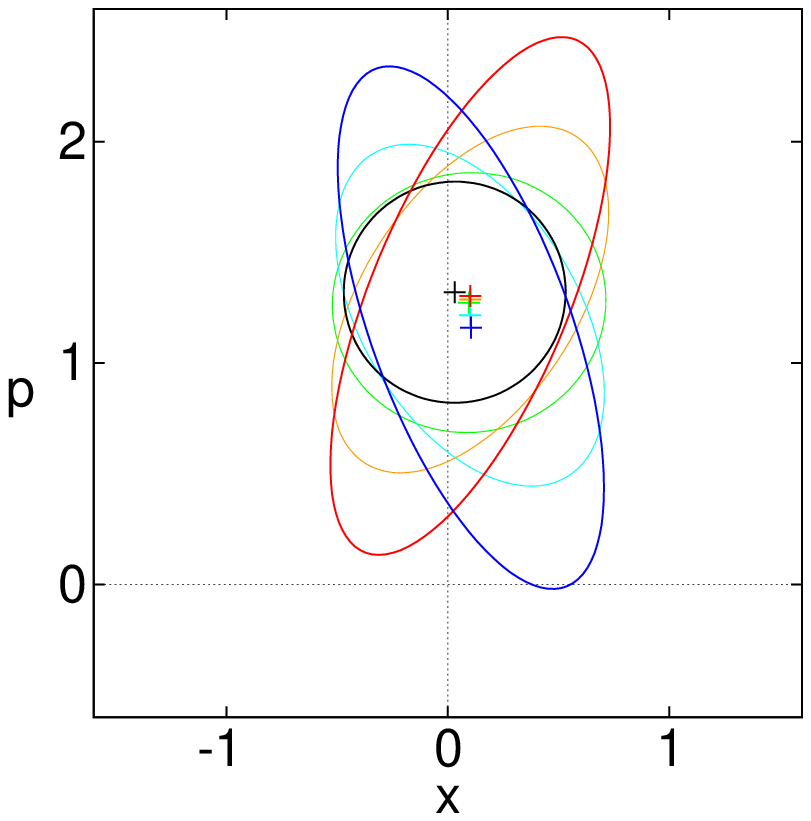}
}
\caption{(color). Input coherent state (black circle) and output states for several $\kappa$.
We assume Gaussian distributions, and show averaged amplitudes, variances.
(a, c, d):~Experimental results for three types of input coherent state $\ket{\alpha}$, where $\alpha$ is the complex amplitude ($\hat{a} = \hat{a}^{(0)} + \alpha$).
(b):~Theoretical prediction with infinite squeezed ancillae for the same input state as (a).}
\label{fig:fit2G}
\end{figure}

In Fig.~3, the elliptic output Wigner functions for $\kappa = 0, \pm 1.0, \pm 2.0$ are shown,
where the position, size, and shape of each ellipse correspond to the averaged amplitudes and variances. 
Fig.~3(a, b) are for the case of $\alpha = x_\In$: (a) experimental results and (b) theoretical, ideal operation.
They agree well in positions and inclinations of ellipses, although the ellipses in Fig.~3(a) are thermalized because of the
finite squeezing of the ancilla states.
We estimate the experimentally obtained $\kappa$ via $\kappa _{\mathrm{act}} = \avg{\hat{p}_\Out}/\avg{\hat{x}_\In}$, and the values obtained are $\kappa _{\mathrm{act}} = 0.00$, $0.95$, $-1.04$, $1.94$, and $-2.02$ for theoretical values $\kappa _{\mathrm{th}}= 0$, $\pm 1.0$, and $\pm 2.0$, respectively.
The differences in inclinations between experimental and ideal Wigner functions are less than $3^\circ$.
The experimental results for the other input states are shown in Fig.~3(c, d).
The change of the amplitude in the input states only affects the positions of the ellipses; the shapes and inclinations of the ellipses remain the same.
We can see in Fig~3(d) that the input amplitude in the $p$ quadrature ($p_\In$) is simply reproduced at the output and is otherwise not affected for any $\kappa$.
All of these results are in good agreement with the theoretical input-output relations.





In Fig.~4(a), the fidelities of the experimental output states compared to the ideal pure output states (i.e. without excess noise) are plotted. The fidelity quantifies the overlap between two quantum states, and it can be calculated as $_\In\bra{\psi}\hatd{U}\hat{\rho}_\Out\hat{U}\ket{\psi}_\In$.
In the case of infinitely squeezed ancillae, unit fidelity is achieved.
In the experiment, excess noises due to finitely squeezed ancillae lead to non-unit fidelities.
Without quantum resources (i.e., using vacuum states for ancillary inputs), the experimental fidelity is $0.62 \pm 0.01$ for $\kappa = 0$, which agrees with the theoretical result $0.63$ derived from Eq.~\eqref{eq:input-output with excess noise}.
With squeezed-vacuum ancillae, the experimental fidelity is $0.81 \pm 0.01$ for $\kappa = 0$, which is much better than the case without nonclassical resources.
For nonzero $\kappa$, the fidelities decrease as $|\kappa |$ increases, because the squeezing of the ideal output state grows compared to that
used in the ancillary states.
Experimental results are very close to the theoretical curves which are calculated from the experimentally obtained squeezing levels of the ancillae.

In Fig.~4(b), the quadrature squeezing of our setup is plotted. Note that the squeezed quadratures are fragile and easily degraded by excess noise.
In the case of infinitely squeezed ancillae, squeezing is obtained for any nonzero $\kappa$; for $\kappa = 0$, on the other hand, the variance of the input coherent state is preserved.
With finitely squeezed ancillae, the excess noises are added to the variances of the ideal outputs.
Without nonclassical resources, squeezing below the SNL is, of course, not obtained for any $\kappa$.
In the case of a squeezing level of the ancillae below -2.9 dB relative to the SNL, the output state is squeezed for sufficiently large $|\kappa |$.
We can observe a noise suppression below the SNL by $0.3 \pm 0.1$~dB for $\kappa = \pm 1.0$, $0.8 \pm 0.1$~dB for $\kappa = \pm 1.5$, and $1.0 \pm 0.1$~dB for $\kappa = \pm 2.0$.

\begin{figure}[tb]
\centering
\subfigure[Fidelities of output states.]{
\includegraphics[clip,scale=0.4]{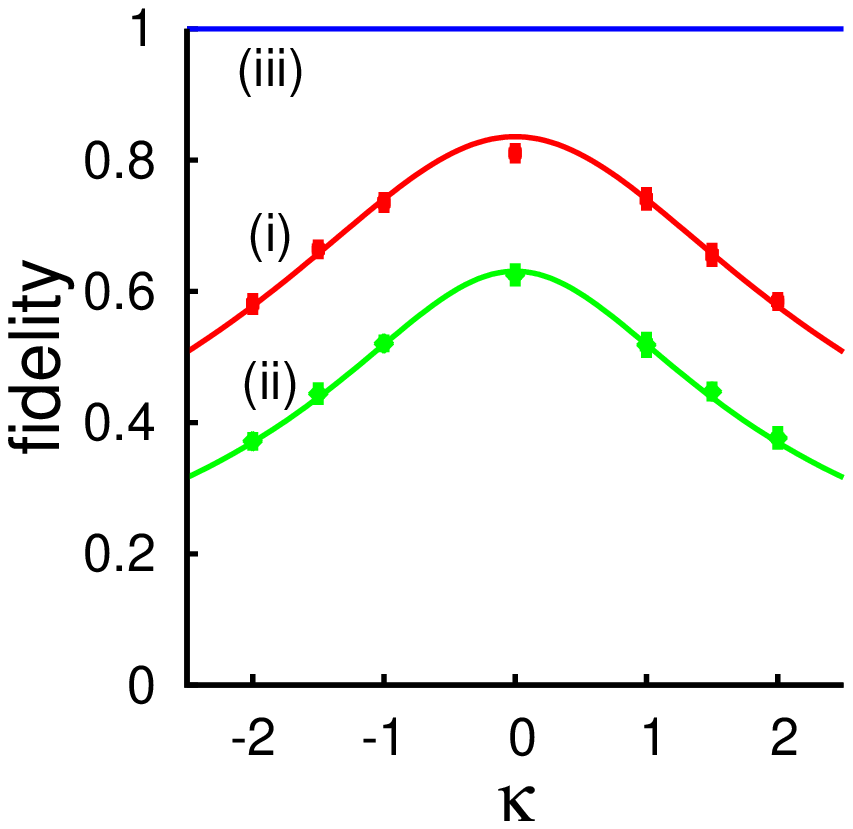}
}
\subfigure[Variances of squeezed quadrature of output states.]{
\includegraphics[clip,scale=0.4]{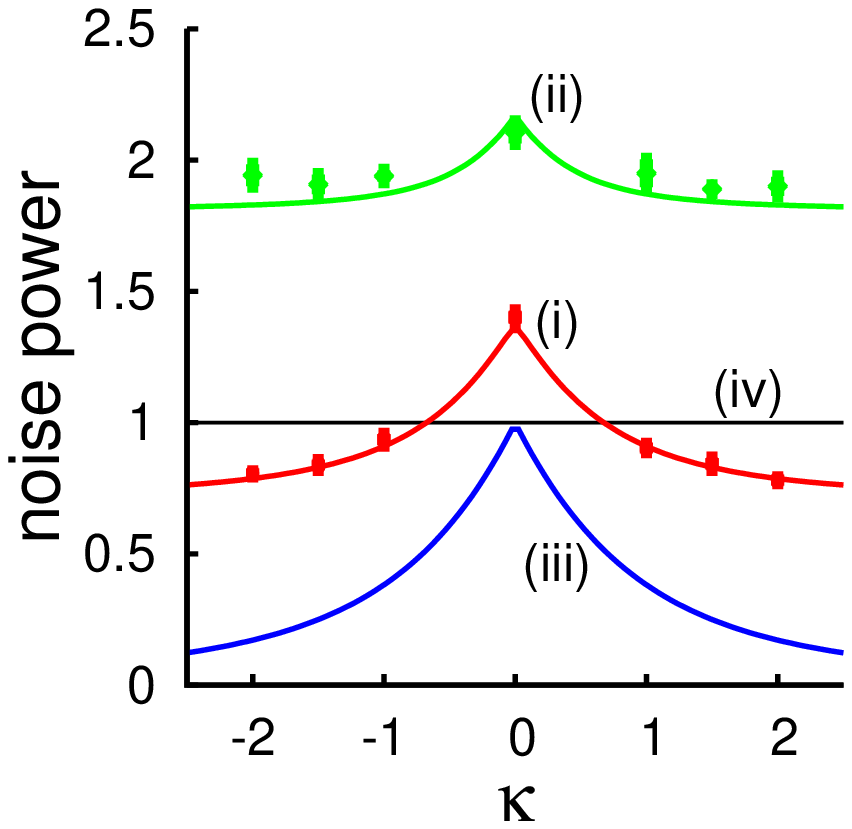}
}
\caption{Fidelities of output states and variances of squeezed quadrature.
(i):~experimental results with squeezed ancillae and their theoretical curves derived from Eq.~\eqref{eq:input-output with excess noise}.
(ii):~experimental results with vacuum-state ancillae and their theoretical curves.
(iii):~theoretical results with infinite squeezing ancillae.
(iv):~shot-noise limit. The vertical axis in (b) is normalized to the shot-noise limit.}
\label{fig:sq_fi}
\end{figure}


In conclusion, we have experimentally demonstrated the canonical quadratic phase gate for CV
in a small cluster computation. The gate is fully controlled by the local oscillator phase
of the homodyne detector. We demonstrated controllability for a set of coherent input states and
we observed sub-shot-noise quadrature variances in the output states, verifying that our measurement-based
gate creates nonclassical states. Concatenating this scheme would enable one to realize any
single-mode Gaussian transformation, efficiently applicable to arbitrary input states including
non-Gaussian states.


This work was partly supported by SCF, GIA, G-COE, and PFN commissioned by the MEXT of Japan, the Research Foundation for Opt-Science and Technology, and SCOPE program of the MIC of Japan.
P.~v.~L. acknowledges support from the
Emmy Noether programme of the DFG in Germany.

\end{document}